\documentclass[aps,prl,reprint,groupedaddres,amsmath,amssymb,floatfix]{revtex4-2}

\usepackage{mhchem}
\usepackage{appendix}
\usepackage{graphicx}
\usepackage{makecell}
\usepackage{color}

\begin{document}
\title{Dirac fermions in the altermagnet Ce$_4$Sb$_3$}
\author{Xue He}
\author{Shihao Zhang}
\email{zhangshh@hnu.edu.cn}
\affiliation{School of Physics and Electronics, Hunan University, Changsha 410082, China}
\begin{abstract}
Altermagnetism exhibits zero net magnetization and spin-splitting electronic structure. The interplay between altermagnetism and topological physics becomes an important topic in the condensed matter physics. In this letter, we propose that the altermagnet Ce$_4$Sb$_3$ hosts Dirac fermions through first-principles calculations and tight-binding model analysis. When subjected to external magnetic field, the altermagnet Ce$_4$Sb$_3$ change from altermagnetic state to ferromagnetic state. Accompanied with magnetic order transition, the four-fold degenerate Dirac points are transited into sextuple degenerate points. Especially, spin-orbital coupling effect has little influence about the Dirac fermions. The topological phases in the altermagnetic state may hold the exotic spin-splitter torque and remarkable nonlinear transport effect. Our work paves the way for exploring the interplay between altermagnetism and non-trivial topological phase.

\end{abstract}
\maketitle

 Recently, altermagnetism has emerged as a significant topic in the condensed matter physics\cite{gRuO2,aCe4Sb3,bCe4Sb3,xiao2023spin,jiang2023enumeration,chen2023spin,ren2023enumeration,gao2023ai,qu2024extremely,tan2024bipolarized,guo2024valley,he2023nonrelativistic,okugawa2018weakly}. The altermagetism exhibits zero net magnetization akin to antiferromagnetism. However, the sublattices with opposite spin can not be interconverted by translation or inversion symmetry, resulting in a broken Kramers degeneracy that gives rise to a spin-split electronic structure. Several materials exhibit this collinear compensated magnetic order, such as RuO$_2$ \cite{aRuO2,bRuO2,cRuO2,dRuO2,eRuO2,fRuO2,gRuO2}, MnF$_2$ \cite{aMnF2,bMnF2,cMnF2}, FeSb$_2$ \cite{FeSb2}, CrSb \cite{CrSb,CrSbexpriment,bCrSb}, $\kappa$-Cl \cite{akCl,bkCl}, Cr$_2$SO \cite{Cr2SO}, RuF$_4$ \cite{aRuF4,bRuF4}, \textcolor{black}{LaMnO$_3$ \cite{okugawa2018weakly}} and V$_2$Se$_2$O \cite{aV2Se2O,bV2Se2O}. In these altermagnets, different sublattices are connected by rotational symmetry, and the energy bands display spin-splitting characteristics even without requiring spin-orbital coupling. Due to their distinctive electronic structure, altermagnetism showcases various novel physical phenomena, including the crystal Hall effect \cite{aHalleffect,bHalleffect,cHalleffect,dHalleffect,eHalleffect,fHalleffect,gHalleffect,samanta2020crystal}, spin currents \cite{aSpincurrents,cSpincurrents,bSpincurrents,dSpincurrents,eSpincurrents,hHalleffect,naka2021perovskite}, crystal thermal transport \cite{aCrystalthermaltransport,hoyer2024spontaneous}, \textcolor{black}{magneto-optical Kerr effect\cite{solovyev1997magneto}} and spin-splitting torque effects \cite{atorque,btorque}.

Moreover, the altermagnetism can coexist with other novel electronic order. The altermagnet LiFe$_2$F$_6$ is theoretically found to host the ferroelectric order and possible spin-triplet exciton state\cite{guo2023altermagnetic}. And the altermagnetism can introduce the finite-momentum Cooper pair and Majornana zero mode in the heterostructure\cite{li2023majorana}. Thus, the interplay between altermagnetism and topological electron has garnered significant public interest. By employing effective models and symmetric analysis, researchers have explored the Weyl nodals and mirror Chern bands in the altermagnet\cite{antonenko2024mirror}. However, the study about the material realization of altermagnet featuring non-trivial topological properties remained scarce yet.

In this letter, we propose that the \textcolor{black}{metallic} altermagnet Ce$_4$Sb$_3$ exhibits the non-trivial topological phase. The first-principles calculations, symmetry analysis and tight-binding model demonstrated that there are Dirac fermions in the altermagnetic state. Under external magnetic field, the altermagnetic order is transited into ferromagnetic order. In the ferromagnetic state, there are sixfold degenerate points protected by $C_{3}$ rotational symmetry. Thus, an external magnetic field can trigger the magnetic order transition, leading to a resultant topological transition. Different from Dirac fermions in the conventional antiferromagnets without spin splitting electronic structures\cite{tang2016dirac}, the Dirac fermions in the altermagnetic state can give rise to enhanced spin-splitter torque and notable nonlinear transport effect.

\begin{figure*}[!htbp]
		\centering
        \includegraphics{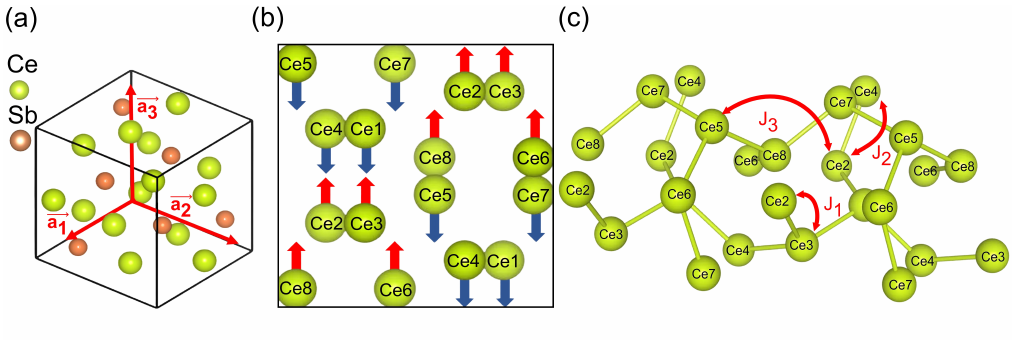}
		\caption{(a) The primitive cell of the \ce{Ce4Sb3} compound. The basis vectors of primitive cell are $\mathbf{a}_1$ = $(-a,a,a)$, $\mathbf{a}_2$ = $(a,-a,a)$, $\mathbf{a}_3$ = $(a,a,-a)$. The lattice constant is $\sqrt{3}a=8.316$\AA{}. (b) The spin configuration of \ce{Ce4Sb3} at 2K in the zero field\cite{nirmala2009understanding,morozkin2011magnetic}. Only Ce atoms are shown and the Ce atomic indices are remarked in the figure. Here we adopt the supercell model for clarity. (c) \textcolor{black}{The illustration about main magnetic exchange couplings. The nearest neighboring exchange coupling is is given by $J_1$ = 2.4781\,meV which prefers to ferromagnetic couplings. And next nearest neighboring exchange interaction whose Ce atoms’ distance is 3.715 \AA{}, is $J_2$ = 0.2217\,meV. Additionally, the exchange interaction at a distance of 8.178\AA{} is given by $J_3$ = -0.1777 meV, which prefers to antiferromagnetic couplings.}}
		\label{fig:1}
\end{figure*}

The \textcolor{black}{metallic} compound \ce{Ce4Sb3} exhibits a cubic structure reminiscent of \ce{Th3P4}-type belonging to the space group $I\bar{4}3d$ (No. 220). Its \textcolor{black}{noncentrosymmetric} unit cell is illustrated in Fig.\ref{fig:1}(a). The primitive cell of \ce{Ce4Sb3} comprises eight Ce and six Sb atoms, whose lattice constant is $\sqrt{3}a=8.316$\AA{} as depicted in Fig.\ref{fig:1}(a). \textcolor{black}{Our Bader charge analysis reveal two species of spins of Ce atoms lost 0.73 and 0.64 electrons per atom, respectively, and six Sb atoms gain 0.9 electrons on average per atom.} Each Ce atom has three nearest neighboring Ce atoms and each atom carries $\sim$ 1\,$\mu_B$ magnetic moment. The magnetic properties of the \ce{Ce4Sb3} compound is unconventional, which is known to exhibit ground-state ferromagnetic (FM) state with possible Kondo interaction\cite{suzuki1990ce4sb3}. However, recent neutron diffraction has unveiled a commensurate antiferromagnetic (AFM) arrangement in bulk \ce{Ce4Sb3} at 2K in the absence of a magnetic field \cite{nirmala2009understanding,morozkin2011magnetic}. The ground magnetic structure of \ce{Ce4Sb3} bulk is shown in the Fig.\ref{fig:1}(b)\cite{nirmala2009understanding,morozkin2011magnetic} and Table.\,S1, whose magnetic space group is No.\,122.336 ($I\bar{4}'2d'$). And the spin space group of altermagnetic spin configuration is $I^{-1}\bar{4}^{-1}2^1d^{\inf m}1$ \cite{chen2023spin,ren2023enumeration}.

Our first-principles calculations reveal that AFM ground state become the ground state while on-site Hubbard parameter of Ce atoms' f orbital is set to $U=1\sim3$ eV as shown in the Table.\,S2. \textcolor{black}{We also performed the calculations about Hubbard parameter with linear response method\cite{PhysRevB.71.035105}, and our results show that self-consistent Hubbard parameter is 3.698\,eV for Ce atoms' f orbitals. The metallic electronic structure provides strong screening effect due to virtual excitation of particle-hole pair in the \ce{Ce4Sb3}, which lead to the small Hubbard value of f-electrons.}

\textcolor{black}{To understand the magnetism of \ce{Ce4Sb3}, we calculated different exchange couplings between Ce atoms as shown in Fig. 1(c). We note that the competition between ferromagnetic coupling $J_2$ and antiferromagnetic coupling $J_3$ may lead to the low-temperature magnetic transition between AFM and FM states.}

\begin{figure*}[!htbp]
		\includegraphics[width=0.9\textwidth]{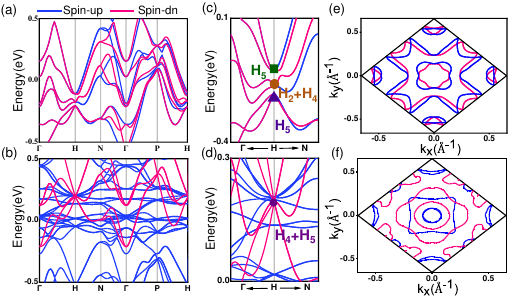} 
		\caption{The calculated electronic band structure and Fermi surfaces of the \ce{Ce4Sb3} compound with $U$=\,2 eV without spin-orbit coupling (SOC) effect. Here the blue lines represent the spin-up bands and the red lines represent the spin-down energy bands. The band structure of the AFM state (a) and the energy bands of the FM state (b) are shown in the figure. Here H point is adopted as $H_y$ point $(0, \pi/a , 0)$. (c) The zoom-in energy bands of AFM state near the Dirac points. (d) The zoom-in electronic structure of FM state near the sextuple degenerate points. \textcolor{black}{The character table of point group $D_{2d}$ is shown in the Table.\,S3}. The $k_x-k_y$ Fermi surfaces of AFM states (e) and FM states (f) are also shown.}
		\label{fig:2}
\end{figure*}

Now we discuss the electronic structure of \ce{Ce4Sb3} bulk. In the nonmagnetic and ferromagnetic cases, the \ce{Ce4Sb3} bulk obeys the $T_d$ point group \textcolor{black}{which lacks inversion symmetry}, whereas in the antiferromagnetic state, the point group is reduced into the $D_{2d}$ point group. The electronic structures of both ferromagnetic (FM) and antiferromagnetic (AFM) cases in the \ce{Ce4Sb3} bulk material are illustrated in the Fig.\ref{fig:2}. In the AFM ground state, the absence of inversion symmetry leads to broken Kramers symmetry, resulting in the altermagnetic energy bands, as depicted in the Fig.\ref{fig:2}(a). These electronic bands exhibit Dirac fermions near the Fermi level at six $H$ points in the entire Brillouin zone as shown in the Fig.\,S2, which reveals the nontrivial topological electronic phases in the AFM state. The Dirac fermions are not sensitive to Hubbard parameters, and they also exist in the altermagnet \ce{Ce4As3} and \ce{Ce4Bi3} as shown in the Fig.\,S3-5. The AFM spin configuration is hold within the primitive cell, thus the band folding affect can be ruled out here. Near the Dirac point located at $H_x$ ($\pm \pi/a$, 0, 0) point, the spin splitting is proportional to $(k_x^2-k_z^2)(k_y-k_z)$, which becomes $\sim (k_y^2-k_z^2)(k_x-k_z)$ near the Dirac point located at $H_y$ (0, $\pm \pi/a$, 0) point. But near the Dirac point located at $H_z$ (0, 0, $\pm \pi/a$) point, the spin splitting becomes $\sim (k_x^2-k_z^2)(k_y^2-k_z^2)$.

We focus on the crucial symmetry protecting the band topology in the altermaget \ce{Ce4Sb3}. Near the Fermi level, the energy bands are mainly contributed by d orbitals of Ce atoms as shown in the Fig.\,S6. Along the $\Gamma - H$ direction, there exists a complex symmetry $\{C_{2y}\tau _x | \mathcal{T}\}$ where $C_{2y}$, $\tau _x$ and $\mathcal{T}$ refer to two-fold rotational operation, sliding operation and time reversal operation, respectively. Under $\{C_{2y}\tau _x | \mathcal{T}\}$ symmetric operation, $(x,y,z)$ becomes $(-x-1/2,y,-z)$, and the $\{C_{2y}\tau _x | \mathcal{T}\}$ symmetry holds the degeneracy between energy bands of different spins. As for the high-symmetric reciprocal path along $H-N$ direction, the symmetry becomes $\{M_{yz}C_{4z}\tau ^\prime\}$. Here $M_{yz}$ and $C_{4z}$ represents the mirror operation and four-fold rotational operation, respectively. And $\tau ^\prime = (-1/4,1/4,-1/4)$ describes the sliding operation. Under $\{M_{yz}C_{4z}\tau ^\prime\}$ operation, $(x,y,z)$ becomes \textcolor{black}{$(-y-1/4,-x+1/4,z-1/4)$}. As shown by detailed analysis in the supplementary materials, the $\{M_{yz}C_{4z}\tau ^\prime\}$ symmetry hold the double orbital degeneracy but lifts the spin degeneracy in the $H-N$ direction. Thus, the Dirac point in the altermagnetic state shown in the Fig.2 is protected by $\{M_{yz}C_{4z}\tau ^\prime\}$ and $\{C_{2y}\tau _x | \mathcal{T}\}$ symmetry.



Under external magnetic field, the \ce{Ce4Sb3} bulk undergoes a magnetic order transition, and ferromagnetic state becomes its ground state. The spin-up energy bands near the Fermi level mainly originates from f-orbitals of Ce atoms. These electrons experience strong interaction, potentially leading to a significant Kondo effect observed in previous experiments. On the other hand, the spin-down electronic structure is mainly contributed by the d-orbitals of Ce atoms. Due to the $C_3$ three-fold rotational symmetry, the energy bands exhibit sixfold degenerate points with high dispersion in the same spin channel, as shown in Fig.\ref{fig:2}(b). Thus, the magnetic field can transit the altermagnetic state into the ferromagnetic state with high topological charge, which may trigger a noteworthy circular photogalvanic effect.

To gain insights into the topological electronic structure, we construct a 16-band tight-binding model based on the eight Ce atom considering spin degree. Each Ce atom possesses an on-site energy of $E_0$, with hopping parameter of $t$ between adjacent Ce atoms. In the paramagnetic phase, the model exhibits a 12-fold degenerate solution at $E_0+t$ and a 4-fold degenerate solution at $E_0-3t$ at the H points. However, in the magnetic phase, Zeeman splitting $\pm m$ arises in different Ce sublattices. In the ferromagnetic phase, the solutions at the H point manifest as two 6-fold degenerate solutions $E_0+t\pm m$ and two 2-fold degenerate solutions $E_0-3t\pm m$. In the antiferromagnetic phase, the solutions evolve into four Dirac points as 
\begin{equation}
   \begin{split}
    E_1&=E_0-t+\sqrt{m^2+4t^2} {\,\,\,\,\rm {(4-fold)}}\\
    E_2&=E_0-t-\sqrt{m^2+4t^2} {\,\,\,\,\rm {(4-fold)}}\\
    E_3&=E_0+t+m {\,\,\,\,\rm {(4-fold)}}\\
    E_4&=E_0+t-m {\,\,\,\,\rm {(4-fold)}}\\
    \end{split}
\end{equation}

Especially, as shown in the Fig.\ref{fig:2}(c), the orbital doublets in the altermagnetic state can be classified into $H_2+H_4$ and $H_5$ representations. The wavefunctions of $H_2+H_4$ representation keep unchanged under $C_2$ operation, but those of $H_5$ representation will flip their phase under $C_2$ operation. 

We also show the Fermi surface of AFM and FM states in the Fig.\ref{fig:2}(e,f). The FM Fermi surface exhibits the highly spin-polarization. But the AFM Fermi surface in the $k_x-k_y$ plane holds the d-wave magnetism without relativistic effect, \textcolor{black}{where spin splitting is proportional to $\cos k_x -\cos k_y$}. It may provide the finite-momentum Cooper pair and induce the unconventional proximitized superconductivity.

\begin{figure}[!htbp]
		\centering
        \includegraphics[width=0.5\textwidth]{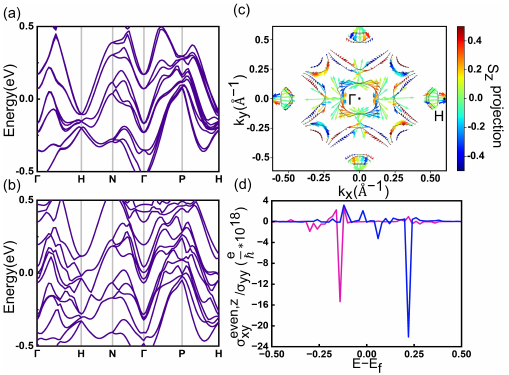}
		\caption{The calculated SOC band structure and spin texture of the \ce{Ce4Sb3} compound with $U$=2\,eV. (a) The band structure of the AFM state. (b) The energy bands of the FM state. (c) The spin texture in the $k_x-k_y$ plane, in which color depicts the spin projection of $S_z$ component. \textcolor{black}{(d) $\mathcal{T}$-even charge-spin conversion ratio $\sigma_{xy}^{\mathrm{even},z}/\sigma$. Here the blue lines and red lines represent the $\sigma_{xy}^{\mathrm{even},z}/\sigma$ in the \ce{CrSb}  \cite{li2024topological} and that in the \ce{Ce4Sb3} with the SOC effect, respectively. The Weyl fermions in the \ce{CrSb} is located near the 0.25\,eV.}}
		\label{fig:3}
\end{figure}

Spin-orbital coupling (SOC) effect always play an important role in the different systems. \textcolor{black}{For example, when SOC effect is taken into consideration, the electronic structure of some altermagnets \ce{RuF4} is dramatically influenced by SOC effect \cite{aRuF4}.} The calculated bands with SOC effect are present in the Fig.\,\ref{fig:3}. In the FM state, due to the strong hybridization between d-orbitals and f-orbitals of opposite spins, the 6-fold degenerate points is fully destroyed in the FM state. In the AFM state, the SOC effects break the aforementioned $\{M_{yz}C_{4z}\tau ^\prime\}$ and $\{C_{2y}\tau _x | \mathcal{T}\}$ symmetry. \textcolor{black}{There is no remarkable canted magnetic moment in the magnetic ground state, which implies the nonexistence of topological Hall effect}. But the energy splitting near the Dirac point is very slight as shown in the Fig.\,\ref{fig:3}(a). \textcolor{black}{The Dirac point is mainly contributed by $d_{xz}$ and $d_{z^2}$ orbitals, and little $d_{yz}$ orbital contributes to Dirac fermions.} Thus, the Dirac fermions in the altermagnetic state still can be regarded as nearly degenerate electronic structure even under SOC effect, \textcolor{black}{in which weak splitting originates from little weight of $\langle d_{xz}|L_zS_z|d_{yz}\rangle$ SOC matrix element}. Thus, our results prove that the Dirac fermions in the altermagnetic state can be observed in the ARPES experiments.

Near the Dirac point, two energy bands have the in-plane spin texture structure under SOC effect as shown in the Fig.\ref{fig:3}(c). But the other two energy bands exhibit the $k$-dependent out-of-plane magnetization. The novel spin textures of four energy bands are attributed by topological electronic structure. We use a minimal model to describe this spin structure near the Dirac point. The SOC Hamiltonian in the first order of $\mathbf{k}$ can be written as $H=k_xs_x\sigma_0+k_ys_y\sigma_0+k_xs_x\sigma_x+k_ys_y\sigma_x$ where $s$ and $\sigma$ refer to the Pauli matrices in the spin and orbital space. By solving this minimal model, we can obtain that two bands with in-plane spin texture and other two bands with out-of-plane collinear spin configuration. The out-of-plane collinear spin configuration is corrected to partially in-plane spin structure by SOC term of high order in $\mathbf{k}$. The SOC term can provide the non-zero Berry curvature $\mathbf{\Omega}(\mathbf{k})$ near the Dirac points, resulting in the notable Hall current $\mathbf{j} \sim \mathbf{E} \times \mathbf{\Omega}$ in which $\mathbf{E}$ refers to electric field.



Now we propose that the second harmonic generation (SHG) can distinguish between altermagnetic and ferromagnetic phases. The polarization of resultant light in the SHG is defined as $P_i(2\omega)=\chi_{ijk}E_j(\omega)E_k(\omega)$, where $E_j$ and $E_k$ denote the electric field of incident light with frequency $\omega$, and $\chi_{ijk}$ is SHG susceptibility. The SHG susceptibility is related to second order optical conductivity $\chi_{ijk}=i\sigma_{ijk}/2\epsilon _0\omega$. If spin orientation is along $\pm z$ direction, the symmetry analysis indicate that Im$\chi_{xyz}=$ Im$\chi_{xzy}=-$ Im$\chi_{yxz}=-$ Im$\chi_{yzx}$ in the altermagetic state, whereas in the ferromagnetic state, Im$\chi_{xyz}=$ Im$\chi_{xzy}=$ Im$\chi_{yxz}=$ Im$\chi_{yzx}$\cite{gallego2016magndata}. Thus, the magnetic transition and topological transition induced by external magnetic field or defect can be distinguished by SHG probe. 

The altermagnets can provide the non-zero spin-splitter torque effect due to their anisotropic band splittings, which are absent in the conventional antiferromagnets. What do Dirac fermions bring to the altermagnets? The double degenerate bands near the Dirac fermions in the altermagnet contribute to the larger nonrelativistic spin current and enhanced spin-splitter torque compared to other altermagnet without electronic degeneracy. \textcolor{black}{To quantify the charge-spin conversion ratio, we consider the spin conductivity with Kubo formula within constant scattering-rate $\Gamma$ approximation. The spin conductivity $\sigma^a_{bc}$ describes the spin-$a$ current $J^a_b$ along $b$ direction under applied electric field along $c$ direction. It can be split into the $\mathcal{T}$-odd part \cite{PhysRevLett.126.127701},} 
\begin{equation}
	\begin{aligned}
	& \sigma_{bc}^{\mathrm{odd},a} \\
	&=-\frac{e\hbar}{V\pi}\mathrm{Re}\sum_{\mathbf{k},m,n}\frac{\langle u_{n}(\mathbf{k})|\hat{J}_{b}^{a}|u_{m}(\mathbf{k})\rangle\langle u_{m}(\mathbf{k})|\hat{v}_{c}|u_{n}(\mathbf{k})\rangle\Gamma^{2}}{(E_{F}-E_{n}(\mathbf{k}))^{2}+\Gamma^{2})(E_{F}-E_{m}(\mathbf{k}))^{2}+\Gamma^{2})},
	\label{Kubo odd equation}
	\end{aligned}
\end{equation}
\textcolor{black}{and $\mathcal{T}$-even part within the $\Gamma\rightarrow0$ limit,}
\begin{equation}
\begin{aligned}
	& \sigma_{bc}^{\mathrm{even},a} \\
	& =\frac{2e\hbar}{V}\mathrm{Im}\sum_{\mathbf{k},m\neq n}\frac{\langle u_{n}(\mathbf{k})|\hat{J}_{b}^{a}|u_{m}(\mathbf{k})\rangle\langle u_{m}(\mathbf{k})|\hat{v}_{c}|u_{n}(\mathbf{k})\rangle}{(E_{n}(\mathbf{k})-E_{m}(\mathbf{k}))^{2}}.
	\end{aligned}
\end{equation}
\textcolor{black}{Here $E_F$ is Fermi energy and $\hat{v}_{c}$ is velocity operator. Our calculations reveal that $\mathcal{T}$-odd charge-spin conversion ratio of \ce{Ce4Sb3} reaches $\sim$ 10\% as shown in the Fig.\,S7, which is close to those of conventional altermagnets without topologically non-trivial fermions \cite{PhysRevLett.126.127701}. As for $\mathcal{T}$-even charge-spin conversion ratio, $\sigma_{bc}^{\mathrm{even},a}/\sigma \to 0$ in the clean limit in the conventional altermagnet \cite{PhysRevLett.126.127701}. But  $\sigma_{bc}^{\mathrm{even},a}/\sigma$ in the \ce{Ce4Sb3} with Dirac fermions and \ce{CrSb} with Weyl fermions \cite{li2024topological} behave remarkable near the topologically non-trivial fermions as shown in the Fig.3(d).} In this viewpoint, Dirac-fermion electronic structures in the altermagnets merge the advantages of altermagnets and Dirac fermions in the conventional antiferromagnets\cite{tang2016dirac}. As for the nonlinear transport induced by quantum geometry in the altermagnet \cite{fang2023quantum}, the Dirac points denote additional interband contribution in the nonlinear transport.

In this work, we report the non-trivial topological fermions in the altermagnet \ce{Ce4Sb3}. Our first-principles calculations, symmetry analysis and tight-binding model prove the 4-fold Dirac fermions in the altermagetic state. When the altermagnetic states are transited into the ferromagnetic states, the topological transition occurs and 4-fold Dirac fermions are evolved into 6-fold degenerate fermions. The symmetry analysis show that magnetic transition and topological transition can be probed by nonlinear optical experiments. The electronic doublet near the Dirac fermions can contribute to enhanced spin-splitter torque and strong nonlinear transport. Therefore, the \ce{Ce4Sb3} altermagnet serves as a promising platform for studying the interplay between topology and altermagnetism.

\textit{Acknowledge.}  We thank Xiaobing Chen for helpful discussions. This work was supported by the National Natural Science Foundation of China (No. 12304217) and the Fundamental Research Funds for the Central Universities from China.

\bibliography{reference}
\bibliographystyle{apsrev4-2}

\clearpage

\appendix
\renewcommand*{\thefigure}{S\arabic{figure}}
\renewcommand*{\thetable}{S\arabic{table}}
\setcounter{figure}{0}
\setcounter{table}{0}
\section{Supplementary Materials}
\section{Calculation methods}
The first-principles calculations were performed using the Vienna Ab-initio Simulation Package (VASP) based on density functional theory (DFT)\cite{VASP,wang2021vaspkit}. The Perdew-Becke-Ernzerhof (PBE) functional with generalized gradient approximation (GGA)\cite{GGA} is adopt to describe the exchange-correlated potential. The cutoff energy of the plane-wave is 410\,eV, and Brillouin zone is sampled by a 8$\times$8$\times$8 mesh. The irreducible representation of wave-functions are calculated by irvsp\cite{irvsp} code. We calculated different exchange couplings by Green function method as implemented in TB2J package\cite{TB2J}.

\section{The spin-wave Hamiltonian}
Here we only consider the nearest neighboring exchange couplings.
\begin{equation}
	\begin{aligned}
		\hat{H}&=-J_{1}\sum_{<i,j>}\hat{S}_{i}^{A/B}\cdot \hat{S}_{j}^{A/B}-J_{2}\sum_{<i,j>}\hat{S}_{i}^{A/B}\cdot \hat{S}_{j}^{B/A}
	\label{Magnon H}
	\end{aligned}
\end{equation}
where $J_1$ is the same sublattice (A or B) exchange, and $J_2$ is the exchange parameter between the different sublattice. We can write spin operators $\hat{S}_{i}^{A/B}=\hat{S}_{i}^{A/B,x}\hat{\boldsymbol{x}}+\hat{S}_{i}^{A/B,y}\hat{\boldsymbol{y}}+\hat{S}_{i}^{\mathrm{A/B,z}}\hat{\boldsymbol{z}}$ for A/B sublattices in cartesian coordinates. The Holstein-Primakoff (HP) transformation\cite{Holstein-Primakoff} can replace the spin operator in the formula (\ref{Magnon H}) with the boson operator.
\begin{equation}
	\begin{aligned}
		\hat{S}_{i}^{A,+}& =\sqrt{2S-\hat{a}_{i}^{+}\hat{a}_{i}}\hat{a}_{i}, \quad
		\hat{S}_{i}^{A,-} =\sqrt{2S-\hat{a}_{i}^{+}\hat{a}_{i}}\hat{a}_{i}^+ \\
		\hat{S}_{i}^{A,z}& =S-\hat{a}_{i}^{+}\hat{a}_{i},\\
	    \hat{S}_{j}^{B,+}& =\sqrt{2S-\hat{b}_{j}^{+}\hat{b}_{j}}\hat{b}_{j}^+ ,\quad
		\hat{S}_{j}^{B,-} =\sqrt{2S-\hat{b}_{j}^{+}\hat{b}_{j}}\hat{b}_{j} \\
		\hat{S}_{j}^{B,z}& =\hat{b}_{j}^{+}\hat{b}_{j}-S.\\
		\label{HP}
	\end{aligned}
\end{equation}
The $\hat{a}_i^{\dagger}(\hat{b}_j^{\dagger})$ and $\hat{a}_i(\hat{b}_j)$ correspond to the bosonic creation and annihilation operators, respectively. And $\hat{S}^{A/B,\pm}=\hat{S}^{A/B,x}\pm i\hat{S}^{A/B,y}$. In the low-excited state, only few spins are deflected, and the average deviation of each spin is small, so $\sqrt{2S}$ can be used instead of the square root of HP transformation, and the fourth order term of the operator is omitted. Then we can make the Fourier transformation
\begin{equation}
	\begin{aligned}
		\hat{a}_{i}^{\dagger}/\hat{b}_{i}=\frac{1}{\sqrt{N}}\sum_{k}\mathrm{e}^{-\mathrm{i}k\cdot r_{i}}\hat{a}_{k}^{\dagger}/\hat{b}_{k}\\
		\hat{a}_{i}/\hat{b}_{i}^{\dagger}=\frac{1}{\sqrt{N}}\sum_{k}\mathrm{e}^{\mathrm{i}k'\cdot r_{i}}\hat{a}_{k'}/\hat{b}_{k'}^{\dagger}
	\end{aligned}
	\label{Fourier transform}
\end{equation}
Here $\hat{a}_k^{\dagger}(\hat{b}_k^{\dagger})$ and $\hat{a}_k(\hat{b}_k)$ satisfy the bosonic commutation relation  $[\hat{a}_{k},\hat{a}_{k'}^{\dagger}]=[\hat{b}_{k},\hat{b}_{k'}^{\dagger}]=\delta_{kk'}$. Combining Ce atomic coordinate Table(\ref{atomic coordinate}) and formula(\ref{HP}), (\ref{Fourier transform}), we get momentum-space Hamiltonian.
Using $\Psi_{+}=(\hat{a}_{k}^{+},\hat{b}_{k})$  or $\Psi_{-}=(\hat{b}_{k}^{+},\hat{a}_{k})$ spinor basis, the spin-wave Hamiltonian can be written as
\begin{equation}
\begin{aligned}
	&\hat{H}_{k,\pm}=\begin{pmatrix}A_k\pm\Delta_k&B_k-\mathrm{i}D_k\\B_k+\mathrm{i}D_k&A_k\mp\Delta_k\end{pmatrix}
	\label{MagnonForm}
\end{aligned}
\end{equation}
Here $A_k$, $\Delta_k$, $B_k$ and $D_k$ terms are 
\begin{widetext}
	\begin{equation}
		\begin{aligned}
			&A_k=4J_{1}S-8J_{2}S-2J_{1}S\left[\cos 2axk_x\cos(a+2ax)k_y+\cos(a+2ax)k_x\cos 2axk_y\right]\\
			&\Delta_k=2J_{1}S\left[\sin 2axk_x\sin(a+2ax)k_y+\sin(a+2ax)k_x\sin 2axk_y)\right] \\
			&B_k=-2J_1S\left[\cos 2axk_y\cos(a+2ax)k_z+\cos(a+2ax)k_x\cos 2axk_z+\cos 2axk_z\cos(a+2ax)k_y+\cos(a+2ax)k_z\cos 2axk_x\right] \\
			&D_k=2J_1S\left[\sin(a+2ax)k_z\cos 2axk_y+\sin 2axk_z\cos(a+2ax)k_y+\cos(a+2ax)k_x\sin 2axk_z+\sin(a+2ax)k_z\cos 2axk_x \right].
		\end{aligned}
	\end{equation}
\end{widetext}

The Hamiltonian (\ref{MagnonForm}) can be diagonalized with $T_{k}^{\dagger}H_{k,+}T_{k}=diag(E_{k,\alpha},E_{-k,\beta})$ and and $T_{k}^{\dagger}H_{k,-}T_{k} = diag(E_{k,\beta},E_{-k,\alpha})$ \cite{hoyer2024spontaneous}. The eigenvalues are as follows
\begin{equation}
	\begin{aligned}
	E_{k,\alpha}&=\epsilon_{k}+\Delta_{k},\quad E_{k,\beta}=\epsilon_{k}-\Delta_{k}\\
	&\epsilon_{k}=\sqrt{A_{k}^{2}-B_{k}^{2}-D_{k}^{2}}
	\label{Magnon solution}
	\end{aligned}
\end{equation}

The frequency difference between spin-wave branches is proportional to $\Delta _k$. When $k_x$ or $k_y$ =0, two branches are degenerate as shown in the calculated magnon dispersion in the Fig.\,S1. Because $\Delta _k$ is the odd function of $k_x$ or $k_y$, two magnon branches exhibit chiral splitting when $k_xk_y \neq 0$. But it should be noted that high Gilbert damping present in this metallic antiferromagnet makes it challenging to detect the chiral splitting in experiments.

\subsection{The symmetry protection}

We use $\{C_{2y}\tau_x|\mathcal{T}\}$ to elaborate how the Ce atom is spatially transformed within the unit cell. Firstly, we replace the fractional coordinates of the Ce atom in Table.\,\ref{atomic coordinate} with Cartesian coordinates as follows.
\begin{widetext}
\begin{equation}
\begin{aligned}
			&Ce1:(ax,ax,ax)\xrightarrow{C_{2y}\mathcal{T}}(-ax,ax,-ax)\xrightarrow{\tau_x}Ce3:(a-ax,ax,-ax)\\
			&Ce2:(ax,-ax,a(1-x))\xrightarrow{C_{2y}\mathcal{T}}(-ax,-ax,ax-a)\xrightarrow{\tau_x}Ce4:(-ax,a-ax,ax) \\
			&Ce5:(\frac{a}{2}+ax,\frac{a}{2}+ax,\frac{a}{2}+ax)\xrightarrow{C_{2y}\mathcal{T}}(-\frac{a}{2}-ax,\frac{a}{2}+ax,-\frac{a}{2}-ax)\xrightarrow{\tau_x}Ce8:(\frac{3a}{2}-ax,ax-\frac{a}{2},\frac{a}{2}-ax) \\
			&Ce6:(-\frac{a}{2}+ax,\frac{a}{2}-ax,\frac{3a}{2}-ax)\xrightarrow{C_{2y}\mathcal{T}}(\frac{a}{2}-ax,\frac{a}{2}-ax,ax-\frac{3a}{2})\xrightarrow{\tau_x}Ce7:(\frac{a}{2}-ax,\frac{3a}{2}-ax,-\frac{a}{2}+ax)
\end{aligned}
\label{equation9}
\end{equation}
\end{widetext}

Here we summarize the symmetric operators in the Table.\,S1.
\begin{table}[htbp]
	\centering
	\caption{The fractional coordinates of Ce atoms in the primitive cell. Here the basis vectors of primitive cell are $\mathbf{a}_1$ = $(-a,a,a)$, $\mathbf{a}_2$ = $(a,-a,a)$, $\mathbf{a}_3$ = $(a,a,-a)$. The lattice constant is $\sqrt{3}a=8.316$\AA{}. And the transformations under $M_{yz}C_{4z}\tau ^\prime$ operation and $C_{2y}\tau _x$ operation are listed. For example, the site 1 is transformed into site 7 under $M_{yz}C_{4z}\tau ^\prime$ operation. Under $C_{2y}\tau _x$ operation, the site 1 is transformed into site 3. }
	\begin{tabular}{ccccc}
		\hline\hline
            Atomic index & Fractional coordinates & spin & $M_{yz}C_{4z}\tau ^\prime$ & $C_{2y}\tau _x$\\
		\hline
            Ce1&$(x,x,x)$&$\downarrow$ & Ce7 & Ce3\\
		Ce2&$(\frac{1}{2}-x,\frac{1}{2},0)$&$\uparrow$ & Ce6 & Ce4\\
		Ce3&$(0,\frac{1}{2}-x,\frac{1}{2})$&$\uparrow$ & Ce8 & Ce1\\
		Ce4&$(\frac{1}{2},0,\frac{1}{2}-x)$&$\downarrow$ & Ce5 & Ce2\\
		Ce5&$(\frac{1}{2}+x,\frac{1}{2}+x,\frac{1}{2}+x)$&$\downarrow$ & Ce4 & Ce8\\
		Ce6&$(1-x,\frac{1}{2},0)$&$\uparrow$ & Ce2 & Ce7\\
		Ce7&$(\frac{1}{2},0,1-x)$&$\downarrow$ & Ce1 & Ce6\\
		Ce8&$(0,1-x,\frac{1}{2})$&$\uparrow$ & Ce3 & Ce5\\
		\hline\hline
         \label{atomic coordinate}
	\end{tabular}
\end{table}

Along the $\Gamma$-$H_y$ direction ($k_x=k_z=0$), the wave function obey the symmetry $\{C_{2y}\tau_x|\mathcal{T}\}$. Under symmetric operation $\{C_{2y}\tau_x|\mathcal{T}\}$, $(x,y,z)$ is transited into $(-x-1/2,y,-z)$,and $(k_x,k_y,k_z)$ becomes $(k_x,-k_y,k_z)$. Under $M_{yz}C_{4z}\tau ^\prime$ operation, $(x,y,z)$ becomes $(-y-1/4,-x+1/4,z-1/4)$, and the H-N direction satisfying $k_x+k_y=0$ obeys this symmetric operation. 

\begin{table}[htbp]
	\centering
		\caption{The energy differences $\Delta$E = $E_{AFM}-E_{FM}$ calculated with different U parameters.}
        \setlength{\tabcolsep}{7mm}
		\begin{tabular}{cc}
			\hline\hline
			\text{U}(eV)& $\Delta$E (eV)\\
			\hline
                0& 0.26879\\
                1&-0.71818\\
			2&-2.83711\\
            3&-4.57766\\
			4& 0.25884\\
            5& 0.23083\\
			6& 0.20625\\
            7& 0.18279\\
			\hline\hline
		\end{tabular}
\end{table}

\subsection{The tight-binding model}
Now we use the eight-atom tight-binding model to describe the electronic structure of nonmagnetic state. Each atom has three nearest neighboring sites, so the tight-binding Hamiltonian can be written as
\begin{widetext}
\begin{equation}
    H_0 = \left[\begin{array}{cccccccc}
        E_0 & t_0e^{i\mathbf{k}\cdot\mathbf{r}_{21}} & t_0e^{i\mathbf{k}\cdot\mathbf{r}_{31}} & t_0e^{i\mathbf{k}\cdot\mathbf{r}_{41}} & 0 & 0 & 0 & 0 \\
        t_0e^{i\mathbf{k}\cdot\mathbf{r}_{12}} & E_0 & t_0e^{i\mathbf{k}\cdot\mathbf{r}_{32}} & t_0e^{i\mathbf{k}\cdot\mathbf{r}_{42}} & 0 & 0 & 0 & 0 \\
        t_0e^{i\mathbf{k}\cdot\mathbf{r}_{13}} & t_0e^{i\mathbf{k}\cdot\mathbf{r}_{23}} & E_0 & t_0e^{i\mathbf{k}\cdot\mathbf{r}_{43}} & 0 & 0 & 0 & 0 \\
        t_0e^{i\mathbf{k}\cdot\mathbf{r}_{14}} & t_0e^{i\mathbf{k}\cdot\mathbf{r}_{24}} & t_0e^{i\mathbf{k}\cdot\mathbf{r}_{34}} & E_0 & 0 & 0 & 0 & 0 \\
        0 & 0 & 0 & 0 &  E_0 & t_0e^{i\mathbf{k}\cdot\mathbf{r}_{65}} & t_0e^{i\mathbf{k}\cdot\mathbf{r}_{75}} & t_0e^{i\mathbf{k}\cdot\mathbf{r}_{85}}\\
        0 & 0 & 0 & 0 &  t_0e^{i\mathbf{k}\cdot\mathbf{r}_{56}} & E_0 & t_0e^{i\mathbf{k}\cdot\mathbf{r}_{76}} & t_0e^{i\mathbf{k}\cdot\mathbf{r}_{86}}\\
        0 & 0 & 0 & 0 & t_0e^{i\mathbf{k}\cdot\mathbf{r}_{57}} & t_0e^{i\mathbf{k}\cdot\mathbf{r}_{67}} & E_0 & t_0e^{i\mathbf{k}\cdot\mathbf{r}_{87}}\\
        0 & 0 & 0 & 0 & t_0e^{i\mathbf{k}\cdot\mathbf{r}_{58}} & t_0e^{i\mathbf{k}\cdot\mathbf{r}_{68}} & t_0e^{i\mathbf{k}\cdot\mathbf{r}_{78}} & E_0 \\
    \end{array}\right].
\end{equation}
\end{widetext}

In the ferromagnetic state, the tight-binding model becomes 16-band model as following
\begin{equation}
    H_{FM}=(H_0+H_m) \bigoplus (H_0-H_m)
\end{equation}
Here $H_m$ is the Zeeman term diag$\{m,m,m,m,m,m,m,m\}$. Then we can obtain two sextuple degenerate solutions $E_0+t\pm m$ and two two-fold degenerate solutions $E_0-3t\pm m$ at the H point. 

As for the antiferromagnetic state, the Zeeman term $H_m$ in the tight-binding model becomes $H_m$ = diag$\{m,-m,-m,m,m,-m,m,-m\}$. At the H point, the eigenvalues become
\begin{equation}
   \begin{split}
    E_1&=E_0-t+\sqrt{m^2+4t^2} {\,\,\,\,\rm {(four-fold)}}\\
    E_2&=E_0-t-\sqrt{m^2+4t^2} {\,\,\,\,\rm {(four-fold)}}\\
    E_3&=E_0+t+m {\,\,\,\,\rm {(four-fold)}}\\
    E_4&=E_0+t-m {\,\,\,\,\rm {(four-fold)}}\\
    \end{split}
\end{equation}

\begin{table}[htbp]
	\centering
	\caption{List of the character table of point group $D_{2d}$ by using irvsp code at H point of AFM state.}
	\begin{tabular}{cccccc}
        \hline\hline
\text{representation}&\text{E}&2\text{IC$_4$}&\text{C$_2$}&2\text{C$_2$}&2\text{IC$_2$}\\
		\hline\text{H$_1$}&1&\;1&1&\;1\;&\;1\\
		\text{H$_2$}&1&\;1&1&-1&-1\\
		\text{H$_3$}&1&-1&1&1\;&-1\\
		\text{H$_4$}&1&-1&1&-1&1\\
		\text{H$_5$}&2&\;0&-2&0&0\\
         \hline\hline
	\end{tabular}
\end{table}

\begin{figure*}[!htbp]
\centering
\includegraphics[width=0.4\textwidth]{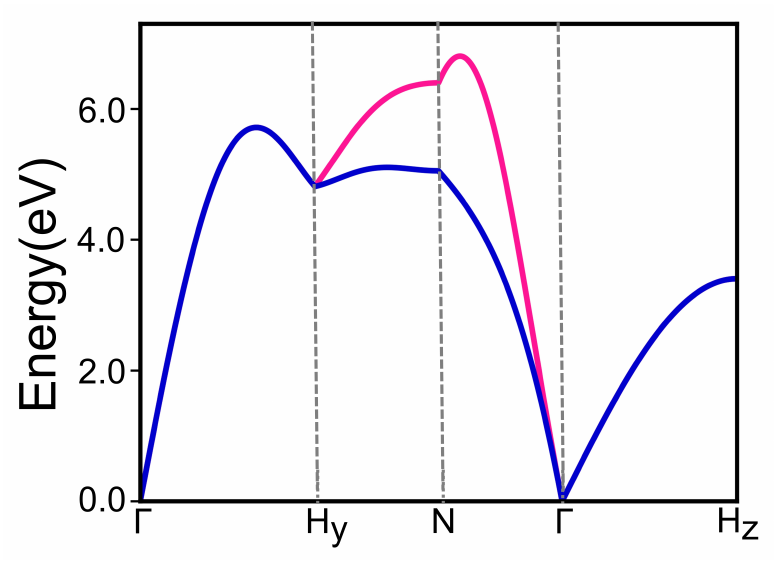}
\caption{The chiral splitting magnon spectrum with only considering nearest neighboring exchange couplings J=-0.43\,eV. The blue (red) line corresponds to different magnon branches. Here $H_y$ and $H_z$ points refer to $(0,\pi/a,0)$ and $(0,0,\pi/a)$, respectively.}
\label{figS0}
\end{figure*}

\begin{figure*}[htbp]
	\centering
	\includegraphics[width=0.8\textwidth]{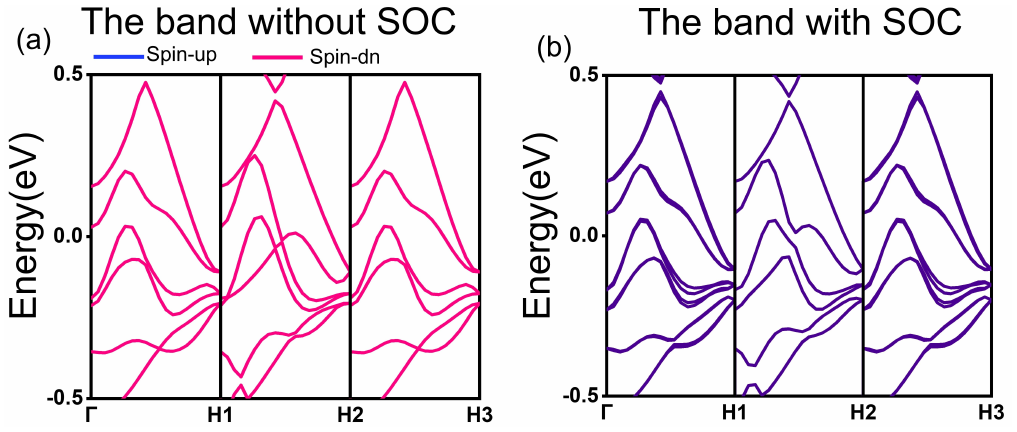}
	\caption{(a) The electronic band structure of the \ce{Ce4Sb3} compound without SOC in the AFM at $U$=\,2eV, where the blue lines represent the spin-up bands and the red lines represent the spin-down bands. (b) The electronic band structure of \ce{Ce4Sb3} with SOC in the AFM at $U$=\,2eV. }
	\label{FigS11}
\end{figure*}
	
\begin{figure*}[htbp]
	\centering
	\includegraphics[width=0.8\textwidth]{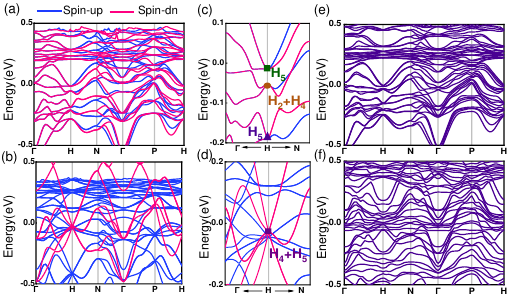}
	\caption{At $U$=\,1 eV, the electronic band structure and irreducible representation of the \ce{Ce4Sb3} compound, where the blue lines represent the spin-up bands and the red lines represent the spin-down bands. (a) Band structure of the AFM state without SOC. (b) Band structure of the FM state without SOC. (c) Irreducible representation of the AFM state at quadruple degenerate points. (d) Irreducible representation of the FM state at sextuple degenerate points. (e) Band structure of the AFM state with SOC. (f) Band structure of the FM state with SOC.}
	\label{FigS1}
\end{figure*}
\begin{figure*}[h]
	\centering
	\includegraphics[width=0.8\textwidth]{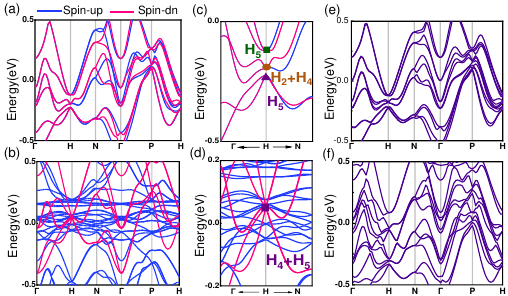}
	\caption{At $U$=\,3 eV, the electronic band structure and irreducible representation of the \ce{Ce4Sb3} compound, where the blue lines represent the spin-up bands and the red lines represent the spin-down bands. (a) Band structure of the AFM state without SOC. (b) Band structure of the FM state without SOC. (c) Irreducible representation of the AFM state at quadruple degenerate points. (d) Irreducible representation of the FM state at sextuple degenerate points. (e) Band structure of the AFM state with SOC. (f) Band structure of the FM state with SOC.}
	\label{FigS2}
\end{figure*}

\begin{figure*}[h]
	\centering
	\includegraphics[width=0.8\textwidth]{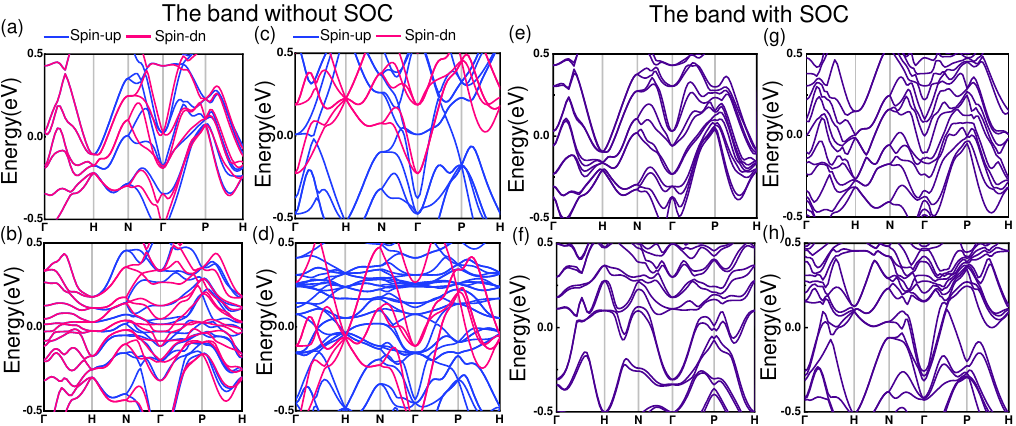}
	\caption{The band structure of the \ce{Ce4Bi3}  compound in the first row and the \ce{Ce4As3} compound in the second row, where the blue line represents the spin up and the red line represents the spin down. (a)-(b) Band structure of AFM state without SOC. (c)-(d) Band structure of FM state without SOC. (e)-(f) Band structure of AFM state with SOC. (g)-(h) Band structure of FM state with SOC.}
	\label{FigS4}
\end{figure*}

\begin{figure*}[!htbp]
\centering
\includegraphics[width=0.8\textwidth]{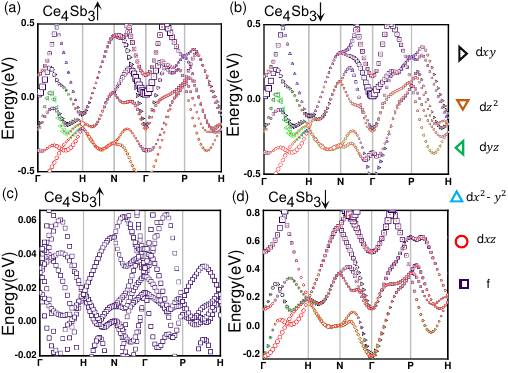}
\caption{At $U$=\,2 eV, orbital resolved energy bands of the \ce{Ce4Sb3} compound without SOC effect. (a) and (b) figures represent the AFM state. (c) and (d) figures refer to the FM state.}
\label{figS3}
\end{figure*}

\begin{figure*}[!htbp]
\centering
\includegraphics[width=0.8\textwidth]{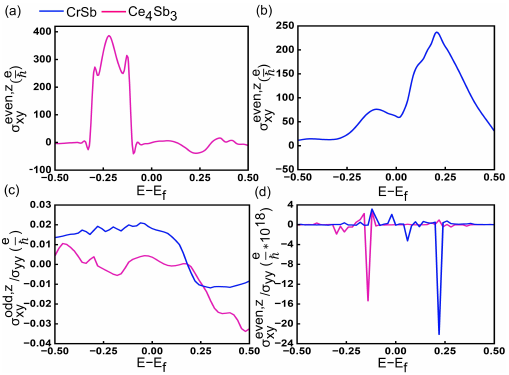}
\caption{The calculated Spin conductivity and spin current angle of the \ce{Ce4Sb3} compound and \ce{CrSb}, where the blue lines represent the \ce{CrSb} and the red lines refer to the \ce{Ce4Sb3} with the SOC effect. (a) $\mathcal{T}$-even spin Hall conductivity of the \ce{Ce4Sb3} compound. (b) $\mathcal{T}$-even spin Hall conductivity of the \ce{CrSb} compound. (c) $\mathcal{T}$-odd charge-spin conversion ratio as a function of the scattering rate. (d) $\mathcal{T}$-even spin charge-spin conversion ratio.}
\label{figS3}
\end{figure*}

\end{document}